\documentstyle[aps,psfig,twocolumn]{revtex}
\begin{document}
\draft \flushbottom
\twocolumn[
\hsize\textwidth\columnwidth\hsize\csname
@twocolumnfalse\endcsname
\title{Temperature dependent change in the symmetry of the order parameter in an electron-doped high-temperature superconductor}
\author{Hamza Balci and R.~L.~Greene}
\address{Center for Superconductivity Research,
Department of Physics, University of Maryland, College~Park,
MD-20742}
\date{\today}
\maketitle \tightenlines \widetext \advance\leftskip by 57pt
\advance\rightskip by 57pt
\begin{abstract}
We present specific heat measurements which show an unexpected
phase transition from d-wave symmetry to s-wave symmetry as the
temperature is reduced in electron-doped
Pr$_{2-x}$Ce$_{x}$CuO$_{4-\delta}$ (PCCO), in both optimal and
over-doped single crystals. The field dependence of electronic
specific heat (C$_{el}$) is linear at T=2K, consistent with s-wave
symmetry, and non-linear, consistent with d-wave symmetry, at
T$\geq$3K. This behavior is most consistent with a phase
transition in the symmetry of the order parameter as the
temperature is changed. Such a phase transition could be an
explanation for the previous controversial results of different
experiments, performed at different temperatures in the
electron-doped cuprates.
\end{abstract}
]\narrowtext \tightenlines

Particle-hole symmetry is an important ingredient of any theory
attempting to explain the mechanism of high temperature
superconductivity. Despite its fundamental importance, this issue
has not been resolved yet, i.e. there is no consensus as to
whether the electron-doped (n-type) and hole-doped (p-type)
cuprates are essentially the same or very different. The symmetry
of the order parameter is one of the important parameters to
compare the n-type cuprates to the p-type family. Unlike the
p-type cuprates, which are now generally considered to have d-wave
symmetry~\cite{kirtley,harlingen}, the pairing (gap) symmetry of
the n-type cuprates continues to remain controversial, despite the
enormous improvement in the sample quality and experimental
techniques since their discovery over 10 years ago. In addition to
experiments that support an s-wave symmetry~\cite{kim} or a d-wave
symmetry~\cite{kokales,tseui,armitage,blumberg,balci2}, there are
a few experiments that suggest a transition from d-wave symmetry
to s-wave symmetry as the doping is increased~\cite{amlan,skinta}.

Specific heat has traditionally been used to study the
superconducting (SC) gap since it is a direct measurement of the
density of states (DOS). The difficulty of detecting the small
T$^2$-type temperature dependence (~5\% of the phonon specific
heat at 4 K in YBCO~\cite{moler}) expected at zero field from a
d-wave superconductor~\cite{kopnin} makes field dependence studies
the preferred method of studying the gap symmetry. In a simple
isotropic s-wave superconductor, the specific heat is linear in
magnetic field for fields larger than H$_{c1}$
(H$>$H$_{c1}$)~\cite{caroli}. For a d-wave superconductor,
C$_{el}$ has a $\sqrt{H}$-type field dependence in the low
temperature ($T<<T_c$) and intermediate field regime
($H_{c1}<<H<<H_{c2}$)~\cite{volovik}.

In our case measuring the field dependence of C$_{el}$ is
particularly advantageous since our PCCO crystals do not have
electronic or nuclear Schottky contributions (see Fig.\
\ref{fig:cvst}) in the temperature (T$>$2K) and field (H$<$10T)
range of our measurements. The field independent addenda leaves
C$_{el}$ the only field dependent part of in our specific heat
measurements. Therefore, our field dependent data was acquired at
a constant temperature while the magnetic field is ramped up (zero
field cooled) or down. No hysteresis has been observed in our
specific heat between increasing and decreasing field.

We measured the heat capacity of
Pr$_{2-x}$Ce$_{x}$CuO$_{4-\delta}$ (PCCO) single crystals grown by
the directional solidification technique with cerium dopings of
x=0.150$\pm$0.005 and 0.165$\pm$0.005. The non-superconducitng
as-grown crystals are annealed at 900 $^o$C (in an inert
atmosphere of flowing argon gas) in order to attain
superconductivity. The cerium concentration of the crystals was
determined using WDX to an accuracy of $\pm0.005$. The experiments
were performed in a Quantum Design PPMS with a modified sample
holder using thermal relaxation calorimetry~\cite{bachmann}.
Au-7$\%$Cu wires (1-3 mil diameter) are used as a weak link to
connect the holder to the thermal bath. The thermometers were
calibrated at different magnetic fields in order to take into
account their small magnetoresistance. Our setup was tested by
measuring a 3 mg high purity copper sample, and a 3.2 mg Nb
sample, and our measurements were  within 5 $\%$ of the standard
values.

Fig.\ \ref{fig:cvst} shows temperature dependence data on an
optimally-doped PCCO crystal with T$_c$=22$\pm$2 K and mass of 3.2
mg in several magnetic fields (H//c-axis). The absence of any
Schottky upturn at low temperatures shows that the sample is free
of any measurable magnetic impurities. The specific heat in this
case can be represented as $C=[\gamma(0)+\gamma(H)]T+\beta T^3$,
where $\gamma(0)$ is the coefficient of the zero field residual
heat capacity, $\gamma(H)$ is the coefficient of the field
dependent C$_{el}$, and $\beta$ is the coefficient of the phonon
specific heat. The gap symmetry is investigated by studying
$\gamma(H)$. For $H\geq H_{c2}$, $\gamma(H)=\gamma_n$, where
$\gamma_n$ is the Sommerfeld constant. The $\gamma(0)$ term is
also found in hole-doped cuprates, and there are several proposals
to explain its origin~\cite{hussey,kresin,phillips2}. By making a
linear fit to the C/T vs T$^2$ data shown in the inset of Fig.\
\ref{fig:cvst} ($\beta$ is kept constant for the two fields), we
find $\gamma(0)$=2.1$\pm$0.2 mJ/moleK$^2$, $\gamma_n$=5.3$\pm$0.2
mJ/moleK$^2$, and $\beta$=0.23$\pm$0.02 mJ/moleK$^4$, which
results in a Debye temperature $\theta_D$=390$\pm$15 K.

The Debye temperature is the same (within the accuracy of our
measurements) in all the crystals we studied, and the value of
$\gamma(0)$ changes between 1.4-2.2 mJ/moleK$^2$ in the different
optimally-doped crystals we studied. The $\gamma_n$ for different
optimally-doped crystals depends on the annealing time (oxygen
content) and we have observed a variation between 4.0-5.3
mJ/moleK$^2$. The upper critical field H$_{c2}$ determined from
specific heat measurements also depends on the annealing time.
Fig.\ \ref{fig:cvst} shows that the upper critical field for this
sample is H$_{c2}<$8T since there is no difference in the specific
heat data between 8T and 10T. The H$_{c2}$ of the different
optimally-doped samples changed between 5.0-7.0 ($\pm$0.5T)
depending on the annealing time of the crystals.

Fig.\ \ref{cvsh-101303b} shows the field dependence of an
optimally-doped crystal (T$_c$=23$\pm$3K, mass=3.4 mg). Only the
field dependent part of C$_{el}$ is shown in Fig.\
\ref{cvsh-101303b} since the zero-field specific heat, which
includes the phonon contribution ($\beta$T$^3$) and the zero field
residual heat capacity ($\gamma$(0)T), is subtracted out. The
field dependence is clearly non-linear at T$\geq$3K. However, a
dramatic change in the field dependence is observed when the
temperature is reduced to 2K, where a linear field dependence
dominates up to about 3T. Fig.\ \ref{cvsh-101303b}-B shows that
the change in the field dependence of C$_{el}$ between 4.5K and 2K
is \textbf{accompanied by a suppression of the electronic specific
heat}. This suggests that the density of electronic states is
suppressed, or possibly \textbf{a gap is opened at the Fermi
level}, since the electronic specific heat is a measure of DOS.

This change in the field dependence of C$_{el}$ from linear at
T=2K to non-linear at higher temperatures has been observed on
four optimally-cerium doped crystals with different oxygen
concentration. In addition to the optimally-cerium doped crystals,
crystals of higher cerium doping also show a similar field
dependence. Fig.\ \ref{fig:cvsh-082803f}-A shows the field
dependence of specific heat for an over-doped crystal
(x=0.165$\pm0.005$, T$_c$=15$\pm$1.5K, mass=2.2 mg). The trend
from linear to non-linear field dependence as the temperature is
increased is also evident in this crystal. In addition, Fig.\
\ref{fig:cvsh-082803f}-B shows that there is a small suppression
of C$_{el}$, and hence electronic DOS, as the temperature is
reduced from 3K to 2K.

There are several possible scenarios to explain such an unusual
change in the field dependence of C$_{el}$: a phase transition in
the symmetry of the order parameter from d-wave at high
temperatures (T$\geq$3K) to s-wave at lower
temperatures(T$\leq$2K); an anisotropic s-wave gap which has a
minimum around 3K; or a vortex-vortex interaction effect which
results in a non-linear field dependence in an s-wave
superconductor at T$\geq$3K. Now we will discuss each of these
scenarios in more detail.

Vortices can interact with each other via quasiparticle (QP)
transfer between their cores. At high magnetic fields (large
number of vortices) or high temperatures (larger vortex-core size)
the quasiparticle wavefunctions in the core of one vortex overlap
with the quasiparticle wavefunctions in neighboring vortices, and
hence inter-vortex quasiparticle transfer becomes possible. These
inter-vortex QP transfers result in a shrinking of the vortex
cores~\cite{sonier-shrink}. The calculations of Ichioka \emph{et
al.}~\cite{ichioka}, which took into account these vortex-lattice
effects, showed that C$_{el}\propto$H$^{0.67}$ at T=0.5T$_c$.
Calculations of Miranovic \emph{et al.}~\cite{miranovic} at lower
temperatures, where less overlap between vortex-cores reduces
these effects and the size of the vortex-core is approximately
constant for H$<$0.5H$_{c2}$, showed that C$_{el}\propto$H for
H$<$0.5H$_{c2}$ at T=0.1Tc.

These ideas were experimentally supported by the non-linear field
dependence of C$_{el}$ observed in some s-wave superconductors,
e.g. V$_3$Si~\cite{ramirez}, NbSe$_2$~\cite{sonier}, and
CeRu$_2$~\cite{hedo}. Despite some complications that were
discovered after the original work, a vortex-lattice
transformation in V$_3$Si ~\cite{sonier-private} and anisotropic
gap in NbSe$_2$~\cite{boaknin-nbsn2}, the fundamental mechanism of
vortex-core shrinking due to vortex-vortex interaction is still
consistent with the non-linear field dependence observed in these
materials~\cite{sonier-private}.

The change in the field dependence of C$_{el}$ in PCCO is in
qualitative agreement with these theories for an s-wave
superconductor. However, in PCCO the change in the field
dependence occurs over a very narrow temperature range
C$_{el}\propto$H$^{0.5}$ at T$\approx$0.2T$_c$ and
C$_{el}\propto$H at T$\approx$0.1T$_c$) compared to the gradual
and slower change based on vortex-lattice effects predicted in
Ref.~\cite{ichioka,miranovic} for a conventional s-wave
superconductor (C$_{el}\propto$H$^{0.67}$ at T=0.5T$_c$ and
C$_{el}\propto$H at T=0.1T$_c$). Hence, it is very unlikely that
the non-linear field dependence we observe in PCCO is due to a
vortex lattice effect. Regardless of the source of the non-linear
field dependence at higher temperatures, the linear field
dependence that we observe at T=2K, which is very similar to what
is predicted by Ref.~\cite{miranovic} for s-wave superconductors,
shows that \textbf{the symmetry of the order parameter is s-wave
at this temperature}.

Another possible explanation for our data is an anisotropic s-wave
gap. However, it is difficult to differentiate between an
anisotropic s-wave superconductor and a d-wave superconductor
using thermodynamic properties because what is usually measured
experimentally is QP excitation energy
$E_k=\sqrt{\xi_k^2+|\Delta_k|^2}$, where $\xi_k$ is the single
particle energy relative to the Fermi level and $\Delta_k$ is the
SC gap, which does not depend on the a phase of the SC gap. In
particular, measurements on YNi$_2$B$_2$C, a highly anisotropic
s-wave superconductor, have shown C$_{el}\propto$H$^{0.5}$ very
similar to d-wave superconductors~\cite{nohara}.

For an anisotropic gap the field dependence could be linear or
non-linear depending on whether the energy of the QP excitations
is below or above the SC gap minimum ($\Delta_{min}$). If we
consider our data in Fig.\ \ref{cvsh-101303b},
$\Delta_{min}\approx3k_B$ ($\approx$0.26 meV), since C$_{el}$ is
linear in H below T=3K, and non-linear at higher temperatures.
However, there is a problem with this picture. In an anisotropic
s-wave gap the magnetic field should have a similar effect on the
QP excitation spectrum as the temperature. The QP's are excited
due to a magnetic field $H$ by an energy
E$_H\approx\Delta_0\sqrt{H/H_{c2}}$~\cite{simonlee}, where
$\Delta_0$ is the gap maximum ($\Delta_0\approx$4meV for
PCCO~\cite{amlan}). At T=3K in Fig.\ \ref{cvsh-101303b}, the field
dependence is non-linear even at fields as low as 0.1 T. Yet at
T=2K the field dependence remains linear up to 3T. Due to the
similar effects of magnetic field and temperature on the QP
excitation spectrum in an anisotropic s-wave gap, for T=2K we
would expect the field dependence to become non-linear at much
lower fields than 3T (H$<$1T). Since this is not what we observe
in our data, we believe that the change from linear field
dependence to non-linear field dependence in C$_{el}$ of PCCO is
\textbf{not} due to an anisotropic s-wave gap.

Finally we consider a phase transition in the symmetry of the
order parameter from s-wave (T$\leq$2K) to d-wave (T$\geq$3K).
This interpretation is particularly attractive since it has the
potential of reconciling the results of many conflicting
experiments. The SQUID, ARPES, and
Raman~\cite{tseui,armitage,blumberg} measurements that suggested
d-wave symmetry in electron-doped cuprates are almost exclusively
performed above 4.2K. On the other hand penetration depth
(0.4K-Ref.~\cite{kim}) and point contact tunnelling spectroscopy
(1.8K-Ref.~\cite{amlan}), are performed at T$<$2K and they are
consistent with s-wave symmetry. The only inconsistent experiment
with this picture is a penetration depth experiment which
suggested d-wave symmetry down to T=0.4K(Ref.~\cite{kokales}).

The above qualitative picture of a phase transition is also
consistent with quantitative analysis both at 4.5K and 2K. The
C$_{el}$ of a d-wave superconductor has a $\sqrt{H}$-type field
dependence given by~\cite{wang-junod}:
% for \left(\frac{TH_{c2}^{1/2}}{T_cH^{1/2}}\right)\ll{1}
\begin{equation} C_{el} = A\sqrt{H}=\gamma_n T
\left(\frac{8}{\pi}\right)^{1/2}\left(\frac{H}{H_{c2}/a^2}
\right)^{1/2}.
\end{equation}
This equation is valid at $T\ll T_c$ and H$_{c1}\ll H \ll H_{c2}$.
Fig.\ \ref{cvsh-101303b}-B shows a d-wave fit given by Eq.(1) to
the 4.5 K data and a linear fit to 2K data. In calculating $A$,
the coefficient of the H$^{1/2}$ term, $\gamma_n$=4.2 mJ/mole
K$^2$, H$_{c2}$=7T, and $a$=0.7 were used. The $\gamma_n$ and
H$_{c2}$ are determined from our specific heat measurements. Using
these parameters a reasonably good fit of our data to
$\gamma(H)=AH^{1/2}$ is obtained at T=4.5K (Fig.\
\ref{cvsh-101303b}-B). An s-wave superconductor on the other hand
has a linear field dependence: $C_{el}=\kappa \gamma_n T
H/H_{c2}(T)$, where $\kappa$ is a geometrical constant of order 1.
Using $\gamma_n$=4.2 mJ/mole K$^2$ and H$_{c2}$= 7 T at T=2 K we
find the theoretical $C_{el}=1.2 \kappa H$. A linear fit to our
data yields $C_{el}=2.0 H$. A comparison with data results in
$\kappa$=1.7, which is a reasonable value for $\kappa$.

Such a phase transition should also be observed in the temperature
dependence of the specific heat. We estimate the magnitude of this
feature to be approximately 5\% of the total heat capacity around
3K (the suppression in Fig.2-b is approximately 5\% of the total
heat capacity), and expect it to be spread over a temperature
range of 1-2K. We were not able to resolve this transition in the
temperature dependence of the specific heat because of our limited
resolution (5\%), however more sensitive measurements using ac
heat capacity are in progress.

The data on the over-doped crystal (shown in Fig.\
\ref{fig:cvsh-082803f}) is also consistent with s-wave theory at
T=2K. Using H$_{c2}$= 3.6 T, $\gamma_n$=2.8 mJ/mole K$^2$
(determined as for the optimally-doped crystal), an s-wave
estimate of 1.6$\kappa$ is obtained for the slope at T=2K. A
linear fit to the data yields 2.2 mJ/mole K T. This results in
$\kappa$=1.4 for the over-doped crystal, which is again a
reasonable value for $\kappa$. The low H$_{c2}$ (H$_{c2}$=2.9±0.2T
at T=3K) makes a comparison with d-wave theory difficult since the
condition H$_{c1}$$\ll$H$\ll$H$_{c2}$ is only satisfied in a very
narrow field range. However, the trend from linear to non-linear
field dependence as the temperature is increased suggests that the
x=0.165 crystal behaves similarly to the x=0.150 crystals.

The phase transition suggested by our data is compatible with a
theoretical model proposed by Khodel \emph{et al.}~\cite{khodel}
which suggests a phase transition in the gap symmetry (from nodal
to fully-gapped) in n-type cuprates based on the location of the
Fermi surface "hot-spots". The consistent picture formed by this
model and our data implies that the hot spots and hence the
antiferromagnetic spin fluctuations are an important ingredient of
superconductivity in the n-type cuprates.

In conclusion, the results of our specific heat measurements can
be summarized as:
\begin{itemize}
    \item a dramatic change is observed in the magnetic field dependence of the
    electronic specific heat in a very narrow temperature range; non-linear
    at T$\geq$3K and linear at T=2K.
    \item there is a suppression of DOS, consistent with
opening of a gap at the Fermi surface, as the
    field dependence changes from non-linear to linear.
    \end{itemize}
In light of these observations we conclude that the symmetry of
the order parameter in the n-type cuprates is not a simple d-wave
symmetry as in the p-type compounds. In particular the s-wave
symmetry observed at T=2K is a definite evidence for this
difference. The change in the field dependence of C$_{el}$ is most
consistent with a phase transition in the symmetry of the order
parameter from d-wave to s-wave. Such a phase transition in the
symmetry of the order parameter can explain most of the prior
conflicting symmetry experiments, since almost all experiments
that suggest d-wave symmetry are performed at T$\geq$4K, and those
that suggest s-wave symmetry are performed at T$\leq$2K.

\textbf{Acknowledgements}\newline We would like to thank Victor
Yakovenko, Jeff Sonier, Chris J. Lobb, and Yoram Dagan for helpful
discussions. This work was supported by the NSF DMR 01-02350.

\begin{figure}
\centerline{\psfig{figure=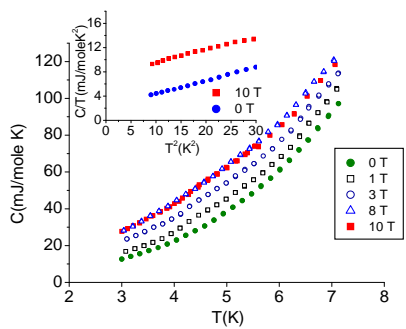,width=6.0cm,height=5.0cm,clip=}}
\caption{Temperature dependence of specific heat for an
optimally-doped crystal, Pr$_{1.85}$Ce$_{0.15}$CuO$_4$
(T$_c$=22$\pm$2K), at different magnetic fields (H$//$c-axis). The
saturation of the specific heat, i.e. field-independent, at 8T
suggests that H$_{c2}\leq$8T. The inset shows C/T vs T$^2$ for 0T
and 10T magnetic fields from which $\gamma(0)$, $\gamma_n$, and
$\beta$ are extracted.}\label{fig:cvst}
\end{figure}

\begin{figure}
\centerline{\psfig{figure=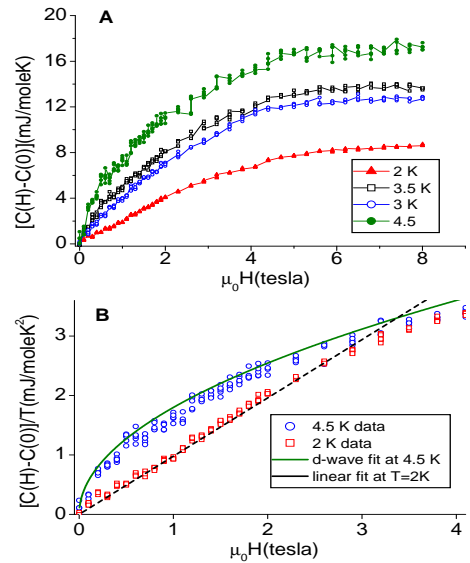,width=6.5cm,height=8.0cm,clip=}}
\caption{\textbf{A-}Field dependent part of the electronic
specific heat (C(H)-C(0)) for an optimally-doped PCCO crystal
(T$_c$=23$\pm$3K) between 2-4.5K. At 2K the field dependence is
linear consistent with a fully-gapped order parameter. At slightly
higher temperatures the field dependence changes from linear to
non-linear. \textbf{B-}The field dependent specific heat is
divided by temperature to show the suppression in the specific
heat, and hence DOS. The curved line is a d-wave fit to 4.5K, and
the dashed line is an s-wave fit to 2K data.}\label{cvsh-101303b}
\end{figure}

\begin{figure}
\centerline{\psfig{figure=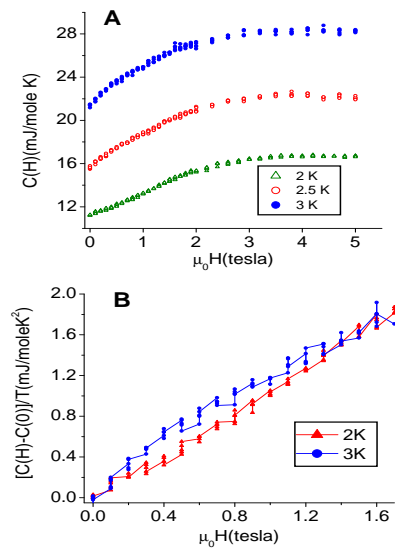,width=6.0cm,height=8.0cm,clip=}}
\caption{\textbf{A}- Field dependence of specific heat for an
over-doped PCCO crystal (T$_c$=15$\pm$1.5K). The field dependence
of the specific heat is similar to the optimally-doped crystals,
i.e. linear at 2K and non-linear at higher temperatures.
\textbf{B}-The field dependent specific heat is divided by
temperature to show the suppression in the specific heat, and
hence DOS, similar to the optimally-doped
crystal.}\label{fig:cvsh-082803f}
\end{figure}

\end{document}